\documentstyle[amssymb,12pt]{article}

\begin{document}

\title{G{\" o}del Type  Metrics in Three Dimensions}
\author{ Metin G{\" u}rses\\
{\small Department of Mathematics, Faculty of Sciences}\\
{\small Bilkent University, 06800 Ankara - Turkey}}

\begin{titlepage}
\maketitle
\begin{abstract}
We show that the G{\" o}del type Metrics in three dimensions with
arbitrary two dimensional background space  satisfy the
Einstein-perfect fluid field equations. There exists only one
first order partial differential equation satisfied by the
components of fluid's  velocity vector field. We then show that
the same metrics solve the field equations of the topologically
massive gravity where the two dimensional background geometry is a
space of constant negative Gaussian curvature. We discuss the
possibility that the G{\" o}del Type Metrics to solve the Ricci
and Cotton flow equations. When the vector field $u^{\mu}$ is a
Killing vector field we finally show that the stationary G{\"
o}del Type Metrics solve the field equations of the most possible
gravitational field equations where the interaction lagrangian is
an arbitrary function of  the electromagnetic field and the
curvature tensors.

\end{abstract}
\end{titlepage}
%\pacs{04.20.Jb, 04.40.Nr}

%\email{gurses@fen.bilkent.edu.tr}
 \maketitle

\section{Introduction}

~~~~ In three dimensions there are several attempts to find exact
solutions of Einstein  and Topologically Massive Gravity field
equations \cite{gur4}-\cite{bar2}. In these efforts authors
usually start with a specific ansatz for the spacetime metrics. It
seems that the {G\" o}del type metrics \cite{gur1}-\cite{gur3}
will be very convenient and practical in searching solutions of
the field equations in three dimensions.

In a $D$ dimensional spacetime the G{\" o}del type metrics are
defined by

\begin{equation}\label{metric}
g_{\mu \nu}=h_{\mu \nu}-u_{\mu}\, u_{\nu}
\end{equation}
where $h_{\mu \nu}$ is the metric of a $D-1$-dimensional locally
Euclidean Einstein space with $h_{\mu \nu}\, u^{\mu}=0$ and
$u_{\mu}$ is a unit timelike vector field with $u^{\mu}=-{1 \over
u_{0}}\, \delta^{\mu}_{0}$. We studied these metrics  when $u_{0}$
is a constant in \cite{gur1} and when $u_{0}$ is not constant in
\cite{gur2}. Although, in these works our approach was independent
of the dimension of the spacetime we focused our attention to the
cases $D > 3$ in much detail. In these works we obtained exact
solutions of various supergravity theories in various dimensions.
 In \cite{gur1} since $u_{0}$ was considered to be a constant our
solutions contain no dilaton field. If  $u_{0}$  is not a constant
it plays the role of the dilaton field. In \cite{gur2} we found
exact solutions of the supergravity theories with dilaton. In
\cite{gur3} we studied the closed timelike curves in  G{\" o}del
type metrics and showed that when the vector field $u_{\mu}$ is
also a Killing vector of the spacetime geometry then there always
exist closed timelike or null curves in G{\" o}del type
spacetimes.

In this work we shall consider the G{\" o}del type metrics in
three dimensions with $u_{0}$ constant (or $g_{00}$ is a
constant). The case with $u_{0}$ is not a constant will be
discussed later. There are several interesting  properties of the
spacetime geometry in three dimensions. In two dimensions since
the Ricci tensor is proportional to the metric then the metrics of
any three dimensional spacetime is of G{\" o}del type. In three
dimensions energy momentum tensor of a Maxwell field of the vector
filed $u_{\mu}$ is equivalent to the energy momentum tensor of a
perfect fluid with stiff equation of state. With these properties
we show in Section 3 that any G{\" o}del type metrics in three
dimensions satisfy the Einstein-Perfect fluid field equations.
Using the result of Section 3 we show in Section 4 that G{\" o}del
type metrics in three dimensions satisfy the field equations of
the topological massive gravity (TMG) provided that the two
dimensional space is a space of constant curvature. We find all
possible G{\" o}del type solutions of TMG and show that our
previous solution \cite{gur4} of TMG  is a special case. The Ricci
and Cotton tensors for the G{\" o}del type metrics take very
simple forms which attracts us to consider the corresponding flow
equations. We study the Ricci and Cotton flow equations in Section
5. In the last Section we construct a closed tensor algebra which
enables the G{\" o}del type metrics to solve the  field equations
of a most general Lagrange function of metric, Ricci, curvature
and the antisymmetric Maxwell tensor field and their covariant
derivatives at all order.

\section{G{\" o}del Type Metrics in General Relativity}

Let $u^{\mu}=-{1 \over u_{0}}\, \delta^{\mu}_{0}$ be a timelike
vector with $u_{0}=$ constant, in $D$ dimensional spacetime $M$
and $h_{\mu \nu}$ be the metric of  $D-1$ dimensional Euclidean
space such that $u^{\mu}\, h_{\mu \nu}=0$. G{\" o}del type of
metrics are defined by \cite{gur1}

\begin{equation}
g_{\mu \nu}=h_{\mu \nu}-u_{\mu}\, u_{\nu}.
\end{equation}
 Let us define an antisymmetric tensor $f_{\mu \nu}$ as

\begin{equation}
f_{\alpha \beta}=u_{\beta, \alpha}-u_{\alpha,\beta}.
\end{equation}
The Christoffel symbol corresponding to the metric (\ref{metric})
is

\begin{equation}
\Gamma^{\mu}_{\alpha \beta}=\gamma^{\mu}_{\alpha \beta}+{1 \over
2}\,(u_{\alpha}\, f^{\mu}\,_{\beta}+u_{\beta}\,
f^{\mu}\,_{\alpha})-{1 \over 2}(u_{\alpha| \beta}+u_{\beta|
\alpha})\,u^{\mu},
\end{equation}
where a vertical stroke denotes covariant derivative with respect
to the Christoffel symbol $\gamma^{\mu}_{\alpha \beta}$ and a
semicolon or nabla $\nabla$ will denote covariant derivative with
respect to the Christoffel symbol $\Gamma^{\mu}_{\alpha \beta}$.
It is easy to show that

\begin{equation}
u^{\alpha}\, \partial_{\alpha}\, u_{\beta}=0,~~~ u^{\alpha}\,
f_{\alpha \beta}=0.
\end{equation}
Then

\begin{equation}
\dot{u}^{\mu}=u^{\alpha}\, u^{\mu}_{\,\,;\, \alpha}=0.
\end{equation}
It is a property of the G{\" o}del type of geometries that the
vector field $u^{\mu}$ is also a timelike Killing vector field of
the spacetime geometry ($M,g$) and hence we have

\begin{equation}\label{killing1}
f_{\alpha \beta}=2 \, u_{\beta ; \, \,\alpha}.
\end{equation}
We define the current vector $j_{\mu}$ corresponding to the
antisymmetric tensor field $f_{\mu \nu}$ as

\begin{equation}
j_{\mu} \equiv f^{\alpha}\,_{\mu \vert \alpha}=\nabla_{\alpha}\,
f^{\alpha}\,\,_{\nu}-{1 \over 2}\, f^2\, u_{\nu},
\end{equation}\label{jey}
where $f^2=f_{\mu \nu}\, f^{\mu \nu}$.

\noindent It is now straightforward to prove the following
Proposition  \cite{gur1}

\vspace{0.3cm}

\noindent {\bf Proposition 1}. {\it Let (M,g) be a stationary
spacetime with the G{\" o}del type metric (\ref{metric}). Let
$h_{\mu \nu}$ be the metric tensor of $D-1$-dimensional locally
Euclidean space, then the Einstein tensor becomes

\begin{eqnarray}
G_{\mu\nu} & = & r_{\mu\nu} - \frac{1}{2} \, h_{\mu\nu} \, r +
\frac{1}{2}\, T^{f}_{\mu \nu} +
\frac{1}{2}\,(j_{\mu}\,u_{\nu}+j_{\nu}\,u_{\mu})+ \nonumber\\
&& \left ( \frac{1}{4}\, f^{2} + \frac{1}{2} \, r \right) u_{\mu}
\, u_{\nu}  - \frac{1}{2}\,(u^{\alpha}\,j_{\alpha})\,g_{\mu \nu}
\, , \label{ein}
\end{eqnarray}
where $T^{f}_{\mu \nu}$ denotes the Maxwell energy-momentum tensor
for $f_{\mu\nu}$, $r_{\mu \nu}$ is the Ricci tensor of
$\gamma^{\mu}\,_{\alpha\beta}$. The Ricci scalar is obtained as
\[ R = r + \frac{1}{4}\, f^2 + u^{\mu} \, j_{\mu} \; , \]
where $r$ denotes the Ricci scalar of $r_{\alpha\beta}$. }

\vspace{0.3cm}

\noindent The above Proposition gives the Einstein tensor of
(\ref{metric}) without any conditions. In order to have a physical
energy momentum distribution we assume that $D-1$-dimensional
space is an Einstein space (a vacuum space with a cosmological
constant) and the current vector field $j_{\mu}$ vanishes
everywhere. Then we have

\vspace{0.3cm}

\noindent {\bf Proposition 2}. {\it Let $(M,g)$ be a stationary
spacetime geometry with the G{\" o}del type metric (\ref{metric}).
Let $h_{\mu \nu}$ be the metric tensor of the $D-1$-dimensional
Einstein space, $r_{\mu \nu}={r \over D-1} \, h_{\mu \nu}$ and let
$j_{\mu}=0$. Then the metric $g_{\mu \nu}$ satisfies the  Einstein
field equations with a charged fluid

\begin{equation}
G_{\mu\nu}  =  \frac{1}{2}\, T^{f}_{\mu \nu} + (p+\rho)\, u_{\mu}
\, u_{\nu} +p g_{\mu \nu}\, , \label{ein}
\end{equation}
with
\begin{eqnarray}
\nabla_{\mu}\, f^{\mu \nu}&=&{1 \over 2}\, f^2 u^{\nu}, \label{eqn06}\\
p&=&{(3-D)\over 2(D-1)}\, r,\\
\rho&=&{1 \over 4} f^2+r.
\end{eqnarray}
Here $p$ is the pressure and $\rho$ is the energy density of the
charged perfect fluid.}

\vspace{0.3cm}

 Here the signs of the fluid pressure and the fluid
energy density depends (in particular for $D \ge  3$) on the sign
of the Ricci scalar of the $D-1$ dimensional Euclidean space with
metric $h_{\mu \nu}$.

\vspace{0.3cm}

\noindent {\bf Corollary 3}. {\it If $h_{\mu \nu }$ is the metric
of a Ricci flat space then the energy momentum distribution for
the Einstein field equations for the metric $g_{\mu \nu}$ becomes
charged dust,i.e.,

\begin{equation}\label{god01}
G_{\mu \nu}={1 \over 2}\,T^{f}_{\mu \nu}+{1 \over 4} f^{2}\,
u_{\mu}\,u_{\nu}
\end{equation}
provided $f$ satisfies the equation

\begin{equation}\label{maxeqn}
f^{\alpha}_{~\,\,\,\beta |\alpha}=0
\end{equation}
and $T^{f}$ is the Maxwell energy momentum tensor for the
antisymmetric tensor $f$
\[
T^{f}_{\mu \nu}=f_{\mu \alpha}\,f_{\nu}\,^{\alpha}-{1 \over 4}\,
f^2\, g_{\mu \nu}
\]
where $f^2=f^{\alpha \beta}\, f_{\alpha \beta}$. Maxwell's
equations (\ref{maxeqn}) can also be written as (\ref{eqn06}) }

\vspace{0.3cm}

Hence G{\" o}del type metrics (\ref{metric}) satisfy the Einstein
field equations with charged dust distributions where the only
field equations are the Maxwell equations (\ref{maxeqn}) or
(\ref{maxeqn1}) and the Ricci flat equations for $h_{\mu \nu}$.
There is no electric field ($u^{\mu}\, f_{\mu i}=f_{0i}=0$), only
the magnetic field exists. We have the gauge freedom

\begin{equation}
{u^{\prime}}_{\mu}=u_{\mu}+\partial_{\mu}\, \Omega
\end{equation}
where $\Omega$ is a function satisfying the condition (recall that
${u^{\prime}}^{\mu}=-{\delta^{\mu}_{0} \over {{u}^{\prime}}_{0}},
~~ u^{\mu}=-{\delta^{\mu}_{0} \over {u}_{0}}$)

\begin{equation}
u^{\prime}_{0}=u_{0}+\partial_{0} \Omega
\end{equation}
Both $u_{0}$ and $u^{\prime}_{0}$ are constants. For the
stationary spacetime , which is the case in this work, we have
$\Omega$ not depending on $x^{0}$ and $u^{\prime}_{0}=u_{0}$
(constant) but this leads to constant $\Omega$.

\begin{equation}
{u^{\prime}}^{\mu}=u^{\mu}+g^{\mu \nu}\,\Omega_{,\nu}
\end{equation}
or

\begin{equation}
{\delta^{\mu}_{0} \over u^{\prime}_{0}}={\delta^{\mu}_{0} \over
u_{0}}-g^{\mu \nu}\,\Omega_{,\nu}
\end{equation}
since $u^{\prime}_{0}=u_{0}$ then $\Omega=constant$. Hence we have
the following Proposition

\vspace{0.3cm}

\noindent {\bf Proposition 4}. {\it The only gauge transformation
\begin{equation}
h^{\prime}_{\mu \nu}=h_{\mu \nu}, ~~~
u^{\prime}_{\mu}=u_{\mu}+\partial_{\mu}\, \Omega
\end{equation}
keeping the stationary  G{\" o}del type metric invariant is the
one with constant $\Omega$.}

\vspace{0.3cm}

In this work we only considered the case where $u_{0}$ is a non
vanishing constant. In \cite{gur2} we have studied the G{\" o}del
type metrics when $u_{0}$ is not a constant. In this case, the
proposed metric yields  exact solutions to the various  theories
with a dilaton field.

\section{G{\" o}del Type Metrics in Three Dimensions}

In three dimensions G{\" o}del type of metrics have very
interesting properties. All three dimensional metrics can be
written as a G{\" o}del type of metric with a non-constant
$u_{0}$.

\vspace{0.3cm}

\noindent {\bf Proposition 5}. {\it  All metrics of the spacetime
geometry are of G{\" o}del type with a non-constant $u_{0}$. Any
three dimensional metric can be written as follows:

\begin{eqnarray}
ds^2&=&-P^2\, (dx^{0})^2+2M dx^{0} dx^{1}+2N dx^{0} dx^{2}+\nonumber\\
&&Q^2 (dx^{1})^2+2L dx^{1} dx^{2}+R^2 (dx^{2})^2 \nonumber\\
&=&-P^2 (dx^{0}-{M \over P^2} dx^{1}-{N \over P^2} dx^{2})^2+
 ({M^2 \over P^2}+Q^2) (dx^{1})^2+ \nonumber\\
&&2(L+{MN \over P^2}) dx^{1} dx^{2} +({N^2 \over P^2}+R^2)
(dx^{2})^2, \label{metric2}
\end{eqnarray}
where $P, M, N, Q, L, R$ are functions of $x^{0}, x^{1}$ and
$x^{2}$. Then the last form  (\ref{metric2}) is of G{\" o}del type
(\ref{metric}) with

\begin{eqnarray}
u_{\mu} dx^{\mu}&=& P\,( dx^{0}-{M \over P^2}\,dx^{1}-{N \over P^2}\, dx^{2}), \nonumber\\
h_{\mu \nu} dx^{\mu } dx^{\nu}&=&({M^2 \over P^2}+Q^2)
(dx^{1})^2+2(L+{MN \over P^2}) dx^{1} dx^{2} +\nonumber \\
&&({N^2\over P^2}+R^2) (dx^{2})^2
\end{eqnarray}
Hence $u_{0}=P$ which is not a constant in general and $h$ is the
metric of a two dimensional locally Euclidean space.}

\vspace{0.3cm}

\noindent {\bf Corollary 6}. {\it  When $(M,g)$ is stationary and
$u_{0}=constant$ then  metric functions depend on $x^{1}$ and
$x^{2}$, and $P=constant$}

\vspace{0.3cm}

\noindent Another interesting property of three dimensions is that
any antisymmetric tensor field $f_{\mu \nu}$ can be expressed as
$\epsilon_{\mu \nu \alpha}\, v^{\alpha}$ where $v^{\alpha}$ is any
vector field. Since $u_{0}$ is constant and $u_{\alpha,0}=0$
(stationarity) then $u^{\alpha}\, f_{\alpha \beta}=0$. This
implies that $v^{\alpha}$ is proportional to $u^{\alpha}$. Hence
we have the following Proposition:

\vspace{0.3cm}

\noindent {\bf Proposition 7}. {\it The antisymmetric tensor
$f_{\mu \nu}$ can be expressed as

\begin{equation}\label{maxeqn3}
f_{\mu \nu}=2 w\, \eta_{\mu \nu \alpha}\, u^{\alpha}
\end{equation}
where $w$ is an arbitrary function and $\eta_{\mu \nu
\alpha}=\sqrt{|g|}\, \epsilon_{\mu \nu \alpha}$. Here
$\epsilon_{\mu \nu \alpha}$ is the totally antisymmetric
Levi-Civita tensor. Hence from (\ref{killing1}) we have

\begin{equation}\label{maxeqn4}
\nabla_{\mu }\, u_{\nu}=w\, \eta_{\mu \nu \alpha}\, u^{\alpha}
\end{equation}
}

Taking the divergence of (\ref{maxeqn3}) we obtain

\begin{equation}\label{maxeqn5}
\nabla_{\mu}\, f^{\mu \nu}={1 \over 2}\, f^2\, u^{\nu}\, +2 w_{,
\alpha}\, \eta^{\alpha \nu \mu}\, u_{\mu},
\end{equation}
where $w^2={1 \over 8}\, f^2$. This leads to the following result:

\vspace{0.3cm}

\noindent {\bf Proposition 8}. {\it The above equations
(\ref{maxeqn5}) imply that in three dimensions the Maxwell
equations (\ref{maxeqn1}) are satisfied if and only if $w$ or
$f^2=constant$. }

\vspace{0.3cm}

\noindent In three dimensions due to the property (\ref{maxeqn3})
the energy momentum tensor of $f_{\mu \nu}$ becomes the energy
momentum tensor of a perfect fluid with $p=\rho$ equation of
state.

\vspace{0.3cm}

\noindent {\bf Proposition 9}. {\it In three dimensions due to the
property (\ref{maxeqn3}) the energy momentum tensor corresponding
to the antisymmetric tensor field $f_{\mu \nu}$ reduces to

\begin{equation}
T^{f}_{\mu \nu}={1 \over 2}\, f^2 u_{\mu}\, u_{\nu}+{1 \over 4}\,
f^2\, g_{\mu \nu},
\end{equation}
where the the energy density and the pressure are respectively
given by

\begin{equation}
\rho={1 \over 4}\, f^2,~~~p={1 \over 4}\, f^2.
\end{equation}
Hence the we have the stiff equation of state $p=\rho$.}

\vspace{0.3cm}

\noindent Then any stationary spacetime metric in three dimensions
with $f^2=$ constant satisfies the Einstein perfect fluid field
equations. We state this as the next proposition which will be
used later for different purposes.

\vspace{0.3cm}

 \noindent {\bf Proposition 10}. {\it Let D=3
in Proposition 2 and use Proposition 9 for the energy momentum
tensor of $f_{\mu \nu}$ then the stationary G{\" o}del type
metrics (\ref{metric}) with constant $f^2$ satisfy the Einstein
field equation with a perfect fluid distribution

\begin{equation}
G_{\mu \nu}= {1 \over 2}\,( f^2 +r_{2})\, u_{\mu}\, u_{\nu}+ {1
\over 8}\, f^2\, g_{\mu \nu},
\end{equation}
where $r_{2}$ is the Ricci scalar corresponding to the two
dimensional metric tensor $h_{\mu \nu}$. Energy density and the
pressure of the fluid are respectively given by

\begin{equation} \label{pres1}
p={1 \over 8}\, f^2, ~~~ \rho={3 \over 8}\, f^2+{1 \over 2}\,
r_{2}
\end{equation}

}

\vspace{0.3cm}

\noindent To have some specific solutions we need a coordinate
chart. For this purpose let us now consider the metric in polar
coordinates (geodesic polar coordinates).

\vspace{0.3cm}

\noindent {\bf Proposition 11}. {\it
 Without loosing any generality we can write the metric
given in (\ref{metric2})in polar coordinates so that  the two
dimensional part the coordinate curves are orthogonal

\begin{equation}\label{metric1}
ds^2=m^2\, dr^2+n^2\, d\theta^2-(u_{0}\,dt+u_{1}\, dr+u_{2}\,
d\theta)^2,
\end{equation}
where $u_{\mu}=(u_{0},u_{1},u_{2})$. Here $u_{0}$ is a constant,
$x^{\mu}=(t,r,\theta)$, the functions $m$, $n$, $u_{1}$ and
$u_{2}$ depend on $r$ and $\theta$. The only field equation
(\ref{maxeqn3}) reduces to a single equation

\begin{equation}\label{maxeqn4}
u_{1, \,\theta}=u_{2, \,r}+2w m\, n,
\end{equation}
which is equivalent to $f^2=8w^2=constant$. As a conclusion, for
any 3 dimensional metric $g_{\mu \nu}$
 with $g_{00}$ is a constant, Eq.(\ref{maxeqn4}) solves
the stationary Einstein-Perfect fluid equations where the pressure
and energy density are given in (\ref{pres1})}

\vspace{0.3cm}

\section{G{\" o}del Type Metrics in Topologically Massive  Theory}

Topologically Massive Gravity (TMG) equations found by Deser,
Jackiw and Templeton (DJT) \cite{djt} with a cosmological constant
are given as follows.

\begin{equation}
G^{\mu}_{\nu}+{1 \over \mu}\,C^{\mu}_{\nu}=\lambda\,
\delta^{\mu}_{\nu}. \label{DJT}
\end{equation}
Here $G_{\mu \nu}$ and $R_{\mu \nu}$ are the Einstein and Ricci
tensors respectively and $C^{\mu}_{\nu}$ is the Cotton tensor
which is given by
\begin{equation}
C^{\mu}_{\nu}=\eta^{\mu \beta \alpha }\, \nabla_{\alpha}\, (R_{\nu
\beta}-{1 \over 4}\,R\, g_{\nu \beta}).
\end{equation}
The constants $\mu$ and $\lambda$ are respectively the DJT
parameter and the cosmological constant.

To solve the DJT field equations some time ago  we have introduced
a method \cite{gur4}. In this method we start with Einstein's
equations with a perfect fluid source

\begin{equation}
G_{\mu \, \nu}=T_{\mu \, \nu} , \label{EIN}
\end{equation}

\noindent with

\begin{equation}
T_{\mu \, \nu}=(p+\rho)\,u_{\mu}\,u_{\nu}+p\,g_{\mu \, \nu},
\end{equation}

\noindent where the fluid equations are obtained through the
conservation equation $\nabla_{\mu}\, T^{\mu \nu}=0$ , $p$ and
$\rho$ are respectively the pressure and energy density of the
fluid and $u_{\mu}$ is the fluid's timelike unit four velocity
vector , i.e, $u^{\mu}\,u_{\mu}=-1$. We have the following result
which was reported  previously \cite{gur4}

\vspace{0.3cm}

\noindent {\bf Proposition 12}: {\it If $p, \rho$ are constants
and

\begin{equation}\label{fluid}
\nabla_{\mu}\, u_{\nu}={\mu \over 3}\, \eta_{\mu \nu
\alpha}\,u^{\alpha}
\end{equation}

\noindent then any solution of the Einstein equations $G_{\mu
\nu}=T_{\mu \nu}$ with a perfect fluid distribution is also a
solution of the TMG (\ref{DJT}) with a cosmological constant
$\lambda={2\,p-\rho \over 3}$ and

\begin{equation}\label{pres2}
p={\mu^{2} \over 9},~~~ \rho={\mu^{2}-9\, \lambda_{0} \over 3}
\end{equation}
where $\lambda_{0}=\lambda+{\mu^{2} \over 27}$. }

\vspace{0.3cm}

If the Ricci scalar $r_{2}$ of the metric $h_{\mu \nu}$ is a
constant then as a consequence of the Proposition 12 any G{\"
o}del type of metrics solve the DJT equations

\vspace{0.3cm}

\noindent {\bf Proposition 13}: {\it Stationary G{\" o}del type
metrics in three dimensions with constant $f^2$, see Proposition
10, solve also the TMG field equations if the two dimensional
background space is of constant Gaussian curvature (or $r_{2}=$
constant) and
\begin{equation}\label{r2}
\mu=3 w,~~~r_{2}=-2(w^2+3 \lambda)
\end{equation}
}

\vspace{0.3cm}

\noindent Eq.(\ref{r2})  implies that the two dimensional geometry
with the metric $h_{\mu \nu}$ is flat if $\lambda_{0}=0$ As an
application of the above Proposition 14 let us consider the
following solution of the TMG  \cite{gur4}

\vspace{0.3cm}

\noindent {\bf Proposition 14}. {\it The following metric solves
the Topologically Massive Gravity equations exactly

\begin{eqnarray}\label{MET}
ds^{2}&=&-a_{0}\, dt^{2}\,+\,2\,q\,dt\,d \theta\,+{-q^{2}+h^{2}\,
\psi \over a_{0}}\, d \theta^{2}\, +\, {1 \over \psi}\, dr^{2} \nonumber \\
&=&-(\sqrt{a_{0}} dt-{q \over \sqrt{a_{0}}}\,d\theta))^2+{h^{2}\,
\psi \over a_{0}}\, d \theta^{2}\, +\, {1 \over \psi}\, dr^{2}
\end{eqnarray}

\noindent where $u_{0}=\sqrt{a_{0}}$, $u_{1}=0$ and $u_{2}=-{q
\over \sqrt{a_{0}}}$ and

\begin{eqnarray*}
\psi&=&b_{0}+{b_{1} \over r^{2}}+{3\, \lambda_{0} \over 4}\,r^{2}\\
q&=&c_{0}+\, {e_{0}\, \mu \over 3}\, r^{2}\\
h&=&e_{0}\,r, ~~~ \lambda_{0}=\lambda+{\mu^{2} \over 27}
\end{eqnarray*}

\noindent where $a_{0}$ , $b_{0}$ , $b_{1}$ , $c_{0}$ and $e_{0}$
are arbitrary constants.}

\vspace{0.3cm} The above solution is very special. All the metric
functions depend only on the radial coordinate $r$.  We may call
this solution as stationary spherically symmetric (solutions not
depending on the angular coordinate $\theta$) G{\" o}del type
metrics.

\vspace{0.3cm}

\noindent {\bf Remark 1}.  In the study of the black holes
solutions in topologically massive gravity, Ait Moussa et al
\cite{clem} considered the solution  of Vuorio \cite{vuor} which
is given by

\begin{equation}
ds^2=-[d \tilde{t}-(2 \cosh \sigma+\tilde{w}) d
\tilde{\varphi}]^2+d\sigma^2+\sinh^2 \sigma\, d \tilde{\varphi}^2.
\end{equation}
where $\tilde{w}$ is a constant. This solution is a special case
of our solution (\ref{MET}) with $\lambda=0, \mu=3$ and

\begin{equation}
b_{0}=-2, b_{1}=0, e_{0}=1/2,a_{0}=1, c_{0}=\tilde{w}-2
\end{equation}
The transformation links our solution to the Vuorio solution is
given as  $t=\tilde{t}, \theta=\tilde{\varphi}, r^2=4(1+\cosh
\sigma)$. Our analysis shows  that the Vuorio solution is also of
G{\" o}del type. Hence one may use a similar analytic continuation
used  in \cite{clem} to our solution (\ref{MET}) with $\lambda \ne
0, \mu \ne 3$ to convert it to a black hole solution of TMG. We
remark \cite{clem2} also that the solution given in (\ref{MET}) is
equivalent to the solution (3.13) of Ait Moussa et al
\cite{clem1}.

\vspace{0.3cm}

Now we will show that, using Proposition 10 it is easy to
generalize the above spherically symmetric solution of TMG. Our
solution (\ref{MET}) given above has  two dimensional space with
metric

\begin{equation}\label{MET2}
ds_{2}^2={h^{2}\, \psi \over a_{0}}\, d \theta^{2}\, +\, {1 \over
\psi}\, dr^{2}
\end{equation}
where the Gaussian curvature  is found as  $K=3
\lambda_{0}=3\lambda+w^2$ which is a constant. Ricci scalar
$r_{2}$  and the Gaussian curvature $K$ are related by $r_{2}=2K$.
To write the above exact solution of the TMG compatible with the
notation in Proposition 10 we get the following identifications
\begin{eqnarray}
m={1 \over \sqrt{\psi}},~~~ n= {h \sqrt{\psi} \over
\sqrt{a_{0}}}\\
u_{1}=0,~~~u_{2}=-{1 \over \sqrt{a_{0}}}(c_{0}+w e_{0} r^2).
\end{eqnarray}
where $\psi$ and $h$ are defined in Proposition 15.

To generalize the above solution (\ref{MET}) we leave the two
dimensional part (\ref{MET2}) the same, i.e., letting

\begin{equation}\label{mn}
m={1 \over \sqrt{\psi}},~~~ n= {h \sqrt{\psi} \over \sqrt{a_{0}}}
\end{equation}
and take the most general solution of Eq.(\ref{maxeqn4}), i.e.,
\begin{equation}
u_{1,\theta}=u_{2,r}+2w {e_{0} \over \sqrt{a_{0}}}r
\end{equation}
We solve $u_{2}$ from this equation as

\begin{equation}\label{u2}
u_{2}=-{1 \over \sqrt{a_{0}}}(c_{0}+w e_{0} r^2)+\int^{r}\, (u_{1,
\theta}) dr
\end{equation}
where $c_{0}$ is arbitrary constant and $u_{1}$ is left free.
Hence the metric

\begin{equation}\label{metric1}
ds^2=m^2\, dr^2+n^2\, d\theta^2-(u_{0}\,dt+u_{1}\, dr+u_{2}\,
d\theta)^2
\end{equation}
with $m$, $n$, and $u_{2}$ given above (\ref{mn}) and (\ref{u2})
respectively, solve the TMG exactly. Here $u_{1}(r, \theta)$ is
left arbitrary which was taken to be zero in our solution
(\ref{MET}).  Hence we have the following result:

\vspace{0.3cm}

\noindent {\bf Proposition 15}. {\it We obtain the most general
stationary solution of Topologically Massive Gravity when $g_{00}$
is a constant. The solution is given in G{\" o}del type where
$u_{0}=\sqrt{a_{0}}$, $u_{1}$ is an arbitrary function of $r$ and
$\theta$, $u_{2}$ is given in (\ref{u2}) and the two dimensional
metric is given in (\ref{MET2}) with constant Gaussian curvature
$K=3 \lambda_{0}$. This solution generalizes our solution
presented in \cite{gur4} }

\section{Ricci and Cotton Flows}

In this section we shall assume that the Gaussian curvature $K$ of
the two dimensional space with metric $h_{\mu \nu}$ is  a
constant. From Proposition 10 we have the Ricci tensor of a
stationary G{\" o}del type metrics

\begin{equation}
R_{\mu \nu}= {1 \over 2}\,( f^2 +r_{2})\, u_{\mu}\, u_{\nu}+ {1
\over 2}\, (r_{2}+{1 \over 2}\, f^2)\, g_{\mu \nu}
\end{equation}
where $r_{2}=2K$ is the Ricci scalar corresponding to the two
dimensional metric tensor $h_{\mu \nu}$. Then we have an exact
solution of the Ricci flow equation

\vspace{0.3cm}

\noindent {\bf Proposition 16}. {\it Let $(M,g)$ be the stationary
G{\" o}del type spacetime with $f^2=8w^2$ constant as in
Proposition 10. Then Ricci flow equation \cite{ham}

\begin{equation}\label{ricci1}
{\partial g_{\mu \nu} \over \partial s}=\xi\, R_{\mu \nu}
\end{equation}
where $s$ is the flow parameter and $\xi$ is an arbitrary
constant, has an exact solution if

\begin{eqnarray}
{\partial u_{\mu} \over \partial s}&=&-\xi \, w^2\, u_{\mu}, \label{flow1}\\
{\partial h_{\mu \nu} \over \partial s}&=&-\xi (K-2\,w^2) \,
h_{\mu \nu}. \label{flow2}
\end{eqnarray}}
The above flow equations (\ref{flow1}) and (\ref{flow2}) are
solved exactly by playing with the constants $b_{0}$, $c_{0}$,
$\lambda_{0}$, and $e_{0}$. As an example $u_{0}$ (or $a_{0}$) has
the following behavior  under the this flow

\begin{equation}\label{u1eqn}
u_{0}=u_{0}(0)\, e^{-\xi w^2\, s}
\end{equation}
where $u_{0}(0)$ is an arbitrary constant. On the other hand
taking the trace of both sides of (\ref{ricci1}) we obtain that
$u_{0}=u_{0}(0)\, e^{-\xi ( w^2+K)\, s}$. Hence comparing with
(\ref{u1eqn}) we get $K=0$.

\vspace{0.3cm}

\noindent {\bf Proposition 17}. {\it The Ricci flow equations have
G{\" o}del type of metrics as exact solutions only when the two
dimensional space is a space of zero curvature}

\vspace{0.3cm}

\noindent {\bf Proposition 18} {\it The Cotton tensor for
stationary G{\" o}del type metrics take the following simple form

\begin{equation}
C^{\mu}\,_{\nu}=-w\,(p+\rho)\, (\delta^{\mu}_{\nu}+3 u^{\mu}\,
u_{\nu}).
\end{equation}}

\vspace{0.3cm}

\noindent Hence the spacetime geometry $(M,g)$ is conformally flat
if $\lambda={\mu^2 \over 9}$. The Cotton flow equation

\begin{equation}\label{cotton1}
{\partial g_{\mu \nu} \over \partial s}=\zeta\, C_{\mu \nu}
\end{equation}
where $\zeta$ is a constant and $s$ is the flow parameter. These
equations were recently used by \cite{bayram}. Here we show that
G{\" o}del type metrics solve exactly the Cotton flow equations
(\ref{cotton1}) only when the Cotton tensor vanishes. For $C_{\mu
\nu}$ given above we have

\vspace{0.3cm}

\noindent {\bf Proposition 19}.{\it Let $(M,g)$ be the stationary
G{\" o}del type spacetime with $f^2=8w^2$ constant as in
Proposition 10. Then Cotton  flow equations (\ref{cotton1}) lead
to the following flow equations for $h_{\mu \nu}$ and $u_{\mu}$

\begin{eqnarray}
{\partial u_{\mu} \over \partial s}&=&-k\, u_{\mu}, \label{flow3}\\
{\partial h_{\mu \nu} \over \partial s}&=&-k \, h_{\mu \nu}.
\label{flow4}
\end{eqnarray}
where $k=\zeta\,w\, (p+\rho)$}. Again the  above flow equations
(\ref{flow3}) and (\ref{flow4}) are solved exactly by playing with
the constants $b_{0}$, $c_{0}$, $\lambda_{0}$, and $e_{0}$. As an
example $u_{0}$ (or $a_{0}$) has the following behavior   under
the Cotton flow

\begin{equation}\label{u2eqn}
u_{0}=u_{0}(0)\, e^{-k s}
\end{equation}
where  $u_{0}(0)$ is an arbitrary constant. On the other hand
taking the trace of both sides of (\ref{cotton1}) we obtain
$u_{0}$ is also constant with respect to the flow parameter $s$.
Comparing this with (\ref{u2eqn}) we obtain $k=0$.

\vspace{0.3cm}

\noindent {\bf Proposition 20}.{\it Let $(M,g)$ be the stationary
G{\" o}del type spacetime with $f^2=8w^2$ constant as in
Proposition 10. Then Cotton  flow equations (\ref{cotton1}) have
exact solutions only when the $k=0$, but this means that the
Cotton tensor vanishes}.

\section{G{\" o}del Type Metrics in Higher Curvature Theories}

In three dimensions when the stationary  G{\" o}del type metrics
with constant $f^2$ have further nice properties. It is possible
to show that the tensors $u_{\mu}$, $f_{\mu \nu}$ and $g_{\mu
\nu}$ satisfy the following tensorial algebra.

\vspace{0.3cm}

\noindent {\bf Proposition 21}. {\it Let D=3 and the metric of
spacetime be  G{\" o}del type with constant $f^2$. Let the two
dimensional space with metric $h_{\mu \nu}$ be a space of constant
Gaussian curvature. Then we have the following closed differential
algebra of the tensors  $u_{\mu}$, $f_{\mu \nu}$ and $g_{\mu \nu}$

\begin{eqnarray}
\nabla_{\mu}\, u_{\alpha}&= &{ 1 \over 2}\, f_{\mu \alpha},\label{alg1} \\
f_{\mu \alpha}&=&w\, \epsilon_{\mu \alpha \sigma} u^{\sigma},\\
\nabla_{\alpha}\, f_{\mu \beta}&=&{1 \over 2} w^2 (g_{\mu
\alpha}\,u_{\beta}-g_{\beta \alpha}\, u_{\mu}),\\
\nabla_{\mu}\,g_{\alpha \beta}&=&0. \label{alg2}
\end{eqnarray}}

\vspace{0.3cm}

From the previous Propositions (in particular Proposition 10) we
can deduce that The Ricci, Einstein , curvature tensors and their
contractions at any order will be the linear sum the tensors
$u_{\mu}\, u_{\nu}$ and $g_{\mu \nu}$. Hence  the tensor
differential algebra introduced in (\ref{alg1})-{\ref{alg2})  is
effective to show that the gravitational field equations, for any
gravitational action. are given as follows

\begin{eqnarray}
G_{\mu \nu}&=&A g_{\mu \nu}+B u_{\mu}\, u_{\nu},\\
\nabla_{\mu}\, f^{\mu \nu}&=&C u^{\nu}
\end{eqnarray}
where $A$, $B$ and $C$ are constants depending upon the theory.
This leads to the following result:

\vspace{0.3cm}

\noindent {\bf Proposition 22}. {\it  Let the action of
gravitation contains all possible combinations of Ricci, curvature
and the antisymmetric tensor $F_{\mu \nu}$ and their covariant
derivatives at any order. Then the tensor differential algebra
introduced in (\ref{alg1})-(\ref{alg2}) is effective to show that
the gravitational field equations are solved when the metric is
the stationary G{\" o}del type metrics with constant $f^2$ and the
two dimensional background is a space of constant Gaussian
curvature $K$, and $F_{\mu \nu}=f_{\mu \nu}$  at all orders of the
string tension parameter. }

\vspace{0.3cm}

\section{Conclusion}

We showed that the metric of any three dimensional stationary
spacetime with $g_{00}$  constant satisfies  the Einstein-perfect
fluid equations. The only differential equation to be solved is a
first order partial differential for the components of the fluid
velocity vector field.  We then showed that in this spacetime
symmetry with $g_{00}$ constant  we find the most general solution
of the TMG. This solution generalizes our previous solution
\cite{gur4} (Proposition 15). We showed that stationary G{\" o}del
type metrics constitute a very simple solution of the Ricci flow
equations an do not solve the Cotton flow equations (Propositions
19 and 20). Finally we discussed the possibility that the
stationary G{\" o}del type metrics form a solution of the low
energy limit of string theory with the most possible interactions
of curvature and antisymmetric field $F_{\mu \nu}$ (Proposition
22).

 \vspace{2cm}
%\newpage
 I would like to thank Professors Gerard Clement and  Atalay Karasu for reading the
 manuscript and constructive comments.
This work is partially supported by the Scientific and
Technological Research Council of Turkey (TUBITAK) and  Turkish
Academy of Sciences (TUBA).


\begin{thebibliography}{EMG}
\bibitem{gur4} M. G{\" u}rses, Class. Quantum Grav. {\bf 11}, 2585 (1994).
\bibitem{clem} K.A. Moussa, G. Clement, and C. Leygnac,
Class. Quantum Grav. {\bf 20},L277-L283(2003).
\bibitem{clem1} K.A. Moussa, G. Clement, H. Guennoune and C. Leygnac,
Phys.Rev. {\bf D78}, 064065 (2008).
\bibitem{bar2} J.D. Barrow, D.J. Shaw, and C.G. Tsagas,  Class. Quantum Grav. {\bf 23},
5291-5322 (2006).
\bibitem{gur1} M. G{\" u}rses, A. Karasu and {\" O}. Sar{\i}o{\~ g}lu,
Class. Quantum Grav. {\bf 22}, 1527-1543 (2005).
\bibitem{gur2} M. G{\" u}rses and {\" O}. Sar{\i}o{\~ g}lu,
Class. Quantum Grav. {\bf 22}, 4699(2005).
\bibitem{gur3} R.J. Gleiser, M. G{\" u}rses, A. Karasu and {\" O}. Sar{\i}o{\~ g}lu,
Class. Quantum Grav. {\bf 23}, 2653 (2006).
\bibitem{djt} S. Deser, R. Jackiw,  and S. Templeton, Phys. Rev.
Lett. {\bf 48}, 975 (1982); Ann. Phys., NY {\bf 140}, 372 (1982).
\bibitem{vuor} I. Vuorio, Phys. Lett. {\bf B163}, 91 (1985); R.
Percacci, P. Sodano, and I. Vuorio, Ann. Phys. NY, {\bf 176}, 344
(1987).
\bibitem{clem2} G. Clement, {\it Private Communication.}
\bibitem{ham} R.S. Hamilton, J. Differ. Geom. {\bf 17}, 255
(1982).
\bibitem{bayram} A. U. Ki\c{s}isel, {\" O}. Sar{\i}o{\~ g}lu, and B.
Tekin, Class. Quantum. Grav. {\bf 25}, 165019 (2008).

\end{thebibliography}
\end{document}